\documentclass[12pt]{article}
\usepackage{epsfig,amsfonts,amssymb}
\usepackage{hyperref}
\usepackage{cite}
\input epsf.sty
\usepackage{cite}

\def\ZZZ{{\hbox{ Z\kern-1.6mm Z}}}
\def\RRR{{\hbox{ R\kern-2.4mm R}}}
\def\CCC{{\hbox{ C\kern-2.0mm C}}}
\def\zzz{{\hbox{z\kern-1mm z}}}

\newcommand{\qeq}{{\hbox{=\kern-2.3mm ? \kern.5mm }}}
\renewcommand{\qeq}{=}

\newcommand{\eps}{\epsilon}

\newcommand{\MM}{{\cal M}}

\newcommand{\wt}{\widetilde}
\newcommand{\wh}{\widehat}

\newcommand{\NN}{{\cal N}}

\newcommand{\be}{\begin{equation}}
\newcommand{\ee}{\end{equation}}
\newcommand{\ben}{\begin{eqnarray}\displaystyle}
\newcommand{\een}{\end{eqnarray}}

\newcommand{\refb}[1]{(\ref{#1})}
\newcommand{\p}{\partial}
\newcommand{\sectiono}[1]{\section{#1}\setcounter{equation}{0}}

\def\one{{\hbox{ 1\kern-.8mm l}}}
\def\zero{{\hbox{ 0\kern-1.5mm 0}}}

\begin{document}

\baselineskip 24pt

\begin{center}
{\Large \bf Two Charge System Revisited: Small Black Holes or
Horizonless Solutions? 
}

\end{center}

\vskip .6cm
\medskip

\vspace*{4.0ex}

\baselineskip=18pt

\centerline{\large \rm   Ashoke Sen }

\vspace*{4.0ex}

\centerline{\large \it Harish-Chandra Research Institute}

\centerline{\large \it  Chhatnag Road, Jhusi, Allahabad 211019, INDIA}

\vspace*{1.0ex}
\centerline{E-mail:  sen at hri.res.in, ashokesen1999 at gmail.com}

\vspace*{5.0ex}

\centerline{\bf Abstract} \bigskip

A two charge system in string theory preserving eight supercharges
can be described as a small black hole that has zero entropy in the
supergravity approximation, but 
classical higher derivative corrections
produce a finite entropy in accordance with the prediction of
microstate counting. 
On the other hand for 
the same system one can construct smooth horizonless
classical solutions
whose geometric quantization 
describes the individual microstates
which contribute to the entropy. In this note we point out 
that there is no duality frame in which
the system admits both these classical
descriptions.
Thus in a given duality frame horizonless 
classical solutions and
small black holes are not alternate descriptions of the same
system; their contributions must be added to get a duality
invariant result for the
macroscopic degeneracy.
We  discuss the significance of this observation for the 
macroscopic computation of the degeneracy of BPS states
for a general system. We also discuss the relationship
between the degeneracy and the index computation and
address the puzzle
regarding absence of small black holes in toroidally
compactified type II string theory.

\vfill \eject

\baselineskip=18pt

\tableofcontents

\sectiono{Introduction and Summary} \label{sintro}

BPS states in string theory carrying only two independent charges
have played an important role in our understanding of black holes
and in particular the effect of higher derivative corrections to the
black hole entropy. An example of this is an elementary heterotic
string carrying $-n$ units of momentum and 
$w$ units of winding charge along a circle\cite{dh1,dh2}.
{}From the counting of the states of a fundamental heterotic string
one knows that the degeneracy of these states grows 
as
\be \label{estat}
d(n,w)\sim
\exp(4\pi\sqrt{nw})\, ,
\ee 
for large $n,w$.
Thus it would be natural to ask whether
this degeneracy can be reproduced by computing the Bekenstein-Hawking
entropy of the BPS black hole carrying the same 
charges\cite{thooft,9309145,9401070,9405117}.
A straightforward analysis shows that in the supergravity limit such
a black hole has zero area event horizon and hence is singular. However
it was shown in \cite{9504147,9506200,9712150} that upon taking 
into account the effect of
higher derivative corrections to the action the black hole could develop a
finite area event horizon; and if this happens then a scaling argument can 
be used to show that the entropy of such a black hole must have the
form $C\sqrt{nw}$ for some numerical constant $C$. Later
the value of $C$ was calculated in vacua with four
non-compact space-time dimensions and found to be $4\pi$, in precise
agreement with the microscopic 
result\cite{0409148,0410076,0411255,0411272,0501014}. 
Explicit computation
of $C$ involved taking into account the effect of a class of higher
derivative corrections in tree level
heterotic string theory using the techniques
developed in 
\cite{9801081,9812082,9904005,9910179,0007195,0009234,0012232}. 

In a related development it was discovered that the microstates describing
the states of the fundamental heterotic string can be represented as
solutions to supergravity equations of motion. 
In the heterotic description
these solutions have source terms 
corresponding to fundamental heterotic
string\cite{9510134,9511053}, but if we use a dual 
description as type IIB string theory then
these solutions are completely 
non-singular\cite{0011217,0012025,0109154,0202072,0212210}. 
The geometric quantization of these solutions leads to a degeneracy
of states that agrees with the microscopic result 
\refb{estat}\cite{0512053}. This seems to suggest an alternative
description of the microstates as smooth classical 
solutions without horizon.\footnote{Such 
smooth classical solutions are sometimes
referred to as fuzzballs, but according to the original definition
fuzzballs include more general configurations, {\it e.g.} classical
solutions with sources (like the fundamental
string source) or quantum states\cite{0007011,0311092,0405017,
0502050,0810.4525}.
In view of this we shall continue to use the phrase `smooth
classical solutions'. 
I wish to thank Samir Mathur for a discussion on
this point. For  reviews on various aspects of
the fuzzball proposal see 
\cite{0701216,0804.0552,0811.0263}.}

The goal of this paper is to reexamine the two charge system carefully
and explore the relationship between these two different 
{\it classical} 
description of the microstates: as small black hole and as
smooth classical solutions.  
In this set up we find that in any given duality frame the
two charge system may have representation either as a small black hole
or as smooth classical solutions, but not both. Thus in a fixed duality
frame black holes and smooth solutions are not alternate descriptions
of the same states; only one of these exists as the possible
classical description
of the microstates. In particular this would imply that
black holes capture information about those states which {\it do
not have representations as smooth horizonless
classical solutions.} 
This result is natural
given that a smooth classical
solution, by definition, takes into account the gravitational backreaction
and yet does not form a horizon. Thus it would be surprising if they 
can also
be represented as black holes in the same duality 
frame.\footnote{Similar remarks can be found {\it e.g.} in
\cite{0005003}.}

Note our emphasis on the word `classical': 
we
are looking for solutions to classical 
equations of motion in string theory without external source terms,
\i.e.\ we include the effect of higher derivative terms in the action but
not string loop corrections. 
Since under
a duality transformation the higher derivative corrections and string loop
corrections mix; our analysis will not be manifestly duality invariant.
The same microscopic configuration may appear as small
black hole in one duality frame and as smooth solutions in another
duality frame.

Even though we have emphasized on working with classical solutions,
this does not mean that we are forced to ignore 
quantum effects. Given a (family of) 
smooth horizonless classical solution(s), we can take
into account quantum effects via geometric quantization\cite{crn}.
On the other hand quantum corrections to a black hole entropy
can be computed
via Euclidean path integral in which the Wick rotated solution
appears as a saddle point. What we do not allow are (black hole)
solutions which come into existence only after inclusion of quantum
corrections to the  effective action {\it computed
in the space-time
background without black holes}. We shall elaborate on this in
\S\ref{small} by showing that if we work in a duality frame
where
classically we have only smooth solutions and no small
black holes, and then try to include (wrongly) the quantum corrections
to the effective action {\it computed in the vacuum},
then in the new description based on the quantum effective
action we may get a small black
hole solution but the smooth solutions will now appear to be
singular.

Typically on the microscopic side we compute an appropriate
index instead of the degeneracy, but on the macroscopic side the
Wald entropy\cite{9307038} and its quantum generalization proposed
in \cite{0809.3304,0903.1477}
measures the
absolute degeneracy. Thus one might wonder whether it is
appropriate to compare the two.
In \S\ref{sindex} we discuss this issue following
\cite{0903.1477}, and show that in some cases 
we can use the result
for degeneracy on the macroscopic side to compute the index
and then compare with the microscopic answer. In this context
we also discuss the puzzle related to the absence of small black
holes in type II string theory representing excited states of the
fundamental type II string.

In section \ref{sprofile} we explore the possibility of comparing
quantities on the macroscopic and the microscopic side which are
not protected by the index theorem. Even though we do not have a
proof that this is impossible, we suggest a mechanism that could
wipe out the information on such quantities.

\sectiono{Small Black Hole or Smooth Solutions?} \label{small}

In this section we shall explore under what condition tree level
higher derivative corrections in a string theory can modify the
near horizon geometry of a two charge system and produce a finite
entropy in agreement with the microscopic result.
The main tool in our analysis will be a limited version of the 
scaling argument used in \cite{9504147}. 
We begin with the observation that
the classical action / equations of motion
of type IIA/IIB string
theory has a scaling symmetry under which 
the dilaton $\phi$ gets shifted by a constant 
$\ln\lambda^{-1}$, all other Neveu-Schwarz-Neveu-Schwarz
(NS-NS) sector fields
remain invariant
and the Ramond-Ramond (RR) 
sector fields are multiplied by $\lambda$. 
The 
effect of this scaling is to multiply the action by $\lambda^2$.
The same symmetry also exists in classical heterotic string theory
except that in this case there are no RR sector fields.
Since magnetic charge is directly related to the magnetic field,
under this symmetry the magnetic charges $\vec q^{~mag}_{NSNS}$ 
for NSNS sector fields 
remain invariant, while the magnetic charges $\vec q^{~mag}_{RR}$ 
for the RR sector
fields scale by $\lambda$. On the other hand since the electric
charge is related to the derivative of the action with respect to the
electric field, the electric charges 
$\vec q^{~el}_{NSNS}$ associated with NSNS sector fields
scale by $\lambda^2$ and the electric charges 
$\vec q^{~el}_{RR}$ associated with the
RR sector fields scale by $\lambda$. 
Finally the black hole entropy, being proportional to the
overall multiplicative factor in the
Lagrangian density, gets scaled by $\lambda^2$. 
This leads to the relation\footnote{Note that this relation does
not require any assumption about the structure of the near horizon
geometry, and holds as long as the $\alpha'$-corrected solution has
a finite Wald entropy. It does however assume that the entropy
for a given set of charges is independent of the asymptotic
value of the dilaton and the moduli arising in the RR sector, 
so that the shift of the dilaton and scaling of the RR fields
at infinity do
not have any effect on the 
entropy. For small black holes in heterotic string
theory, the independence of the near horizon geometry of the
asymptotic values of the dilaton and other moduli was explicitly
proven in \cite{9504147}. If RR field strengths are 
present at infinity, 
as in the case of type IIB on
$AdS_5\times S^5$ with RR 5-form background, then 
our argument
also requires
that
the black hole entropy depends
only on a combination of the asymptotic
RR field strength and the dilaton that
remains invariant under the scaling described here.
In the $AdS_5\times S^5$ example such a combination 
corresponds to the 't Hooft coupling of the dual  $\NN=4$
super Yang-Mills theory.}
\be \label{en1}
S_{BH}\left(\lambda\, \vec q_{RR}, \lambda^2 \, 
\vec q^{~el}_{NSNS}, \vec q^{~mag}_{NSNS}\right)
= \lambda^2\, S_{BH}\left(\vec q_{RR}, 
\vec q^{~el}_{NSNS}, \vec q^{~mag}_{NSNS}\right)\, ,
\ee
in the classical theory.  Here $\vec q_{RR}$ stands for both electric
and magnetic RR charges. In practical terms eq.\refb{en1}
implies that if we have a classical black hole
solution with charges $\left(\vec q_{RR}, 
\vec q^{~el}_{NSNS}, 
\vec q^{~mag}_{NSNS}\right)$ and entropy $S_{BH}$, then
by a simple scaling we can generate another classical solution
with charges $\left(\lambda\, \vec q_{RR}, \lambda^2 \, 
\vec q^{~el}_{NSNS}, \vec q^{~mag}_{NSNS}\right)$ and
entropy $\lambda^2 S_{BH}$.

We shall use this relation to analyze the entropy of two charge
black holes in different descriptions. 
The microscopic system consists of a fundamental heterotic string
in heterotic string theory compactified on $\MM\times S^1$ for
some compact manifold $\MM$, carrying $w$ units of winding and
$-n$ units of momentum along $S^1$. As pointed out in
\S\ref{sintro}, the microscopic entropy of this system is
given by $4\pi\sqrt{nw}$ for large
$nw$. Our goal is to explore the possibility
of realizing this system as a (small)
black hole solution in a classical
string theory.

First consider the heterotic
description where 
both $n$ and $w$ correspond to electric NSNS charges.
Thus the scaling relation \refb{en1} takes the form
\be \label{e3}
S_{BH}(\lambda^2\, n,\lambda^2\, w) = \lambda^2 \, S_{BH}(n,w)\, .
\ee
The microscopic result $4\pi\sqrt{nw}$ is consistent with this
scaling symmetry\cite{9504147}. This shows that in this case
tree level higher derivative corrections could modify the singular
near horizon geometry to correctly reproduce the entropy.
Furthermore explicit computation with
supersymmetrized curvature squared terms in
heterotic string theory vacua with four non-compact
space-time dimensions correctly reproduces the overall coefficient
$4\pi$\cite{0409148,0410076,0411255,0411272,0501014}.
The same result for the entropy is obtained if we include
the Gauss-Bonnet term in the effective action\cite{0505122}. 
General arguments based on $AdS_3/CFT_2$
correspondence shows that this result is exact to all orders in the
derivative expansion\cite{0506176}.\footnote{While for systems
carrying both electric and magnetic charges the 
near horizon geometry can be analyzed 
in a straightforward manner, it is more involved for
a system carrying only electric charges as is the situation here.
See \cite{0611062,0707.3818,0707.4303,0708.0016} 
for detailed discussions on this.
It would be fair to say that
while all approaches lead to the same value $4\pi\sqrt{nw}$ for
the entropy of the small black hole, the detailed structure of the
near horizon geometry still remains a mystery. Given that the
world-sheet theory at the horizon is strongly coupled, this is to be
expected since a field redefinition involving higher derivative
terms can completely change the form of the
near horizon geometry without affecting the entropy.
\label{f2}}
Explicit construction of small black hole
solutions in vacua with five or more non-compact space-time
dimensions has not been performed with
the same degree of certainty. If we include the Gauss-Bonnet term
in the five dimensional action then it does give a non-singular
five dimensional black hole horizon with the correct 
entropy\cite{0505122}, but
it is not clear why we are allowed to ignore the other terms in the
action. Indeed, some recent discussion 
on the subtleties of constructing such solutions in five dimensions
can be found in \cite{0809.4954}.
However in heterotic string
vacua with  five non-compact space-time dimensions
one can explicitly construct 
small black ring solutions describing fundamental heterotic
string with angular momentum 
$J>> \sqrt{nw}$ by exploiting the fact that the local geometry
near the core of a five dimensional small black ring is
identical to that of a four dimensional small
black hole with zero angular 
momentum\cite{0506215, 0511120, 0611166}. 
These reproduce correctly
the microscopic entropy 
\be \label{erotate}
4\pi\sqrt{nw -J}\, . 
\ee
More precisely the
entropy of the small black
ring is given by 
$4\pi\sqrt{nw-QJ}$ where $Q$ is a `dipole 
charge'  describing fundamental heterotic string winding number
along the ring. For given $n,w,J$ the maximum contribution to the
entropy comes
from solutions with $Q=1$. These black rings
are in fact the right objects to
compare with the smooth solutions to be described shortly, since the
latter also describe five dimensional configurations carrying angular
momentum. See footnote \ref{f1}
for a more detailed discussion.

Let us now examine if there are smooth horizonless
classical solutions carrying charges $(n,w)$. 
There are indeed   classical
solutions in heterotic string theory describing these 
microstates\cite{9510134,9511053}, but they
are known to require source
terms and hence are not genuine solutions to the equations of 
motion of classical string field theory.
Thus we conclude that in this description the macroscopic entropy of
the system arises from small black holes / black rings,
and not smooth classical solutions.

Now for $\MM=T^4$, \i.e.\ for
heterotic string theory compactified on a five
dimensional
torus $T^4\times S^1$, 
the same system has a dual description as a
two charge D1-D5 system in type IIB string
theory on $K3\times \wh S^1$
with $w$ denoting the number of D5-branes
wrapped on $K3\times \wh S^1$ and $n$ 
denoting the number of D1-branes
wrapped on $\wh S^1$.
In this case both $n$ and $w$ denote RR charges and the scaling relation
\refb{en1} takes the form
\be \label{e2}
S_{BH}(\lambda\, n , \lambda\, w) = \lambda^2\, 
S_{BH}(n, w)\, .
\ee
This is incompatible
with the microscopic result $4\pi\sqrt{nw}$, 
showing that the classical string theory corrections
cannot reproduce the statistical entropy of the system. 
In fact, had there been a classical black hole solution it would
carry far larger entropy than the microscopic result.
Thus the only consistent scenario is that a (small) black hole
/ black ring
solution carrying these charges does not exist in classical type IIB
string theory.
On the other
hand in this case the microstates of the D1-D5 system can be
described as smooth, horizonless, 
source free solutions to the classical equations
of motion\cite{0109154,0202072,0212210,0704.0690} and 
furthermore the geometric quantization of these
solutions leads to a degeneracy in agreement with 
the microscopic result\cite{0512053}. Thus
we conclude that in this duality frame the two charge system can be
described as smooth solutions but not as small black 
holes or black rings.\footnote{For a distant observer 
the smooth solutions
constructed in \cite{0202072,0212210} look like a 
ring\cite{0105136}
carrying a `dipole charge' of Kaluza-Klein (KK) monopole
associated with the $\wh S^1$ compactification. Under the series
of duality transformations which take a D1-D5 system to the
fundamental heterotic string carrying momentum and winding
along $S^1$, the KK monopole becomes
a fundamental heterotic string extending along the non-compact
directions. This is precisely the structure of
the black ring solution on the heterotic side, carrying
fundamental string winding charge along the ring, besides the
usual momentum and winding charges along $S^1$. Thus the
duality between the solutions in the type IIB and the heterotic
description is also visible locally in five dimensional space-time
as long as we stay far from the core of the ring. 
In general however one should be careful in comparing the
classical profiles / quantum wave-functions of BPS states in
different descriptions since this is not protected against
quantum corrections. This has been discussed briefly in
\S\ref{sprofile}. \label{f1}}

The reader may find our insistence on using classical action
(instead of the quantum effective action) a bit disturbing,
so we shall now explain the difficulties
which arise if we try to use quantum corrected effective action
{\it computed in the vacuum}.
This discussion will be closely related to the one in \cite{0412133}
with a slightly different emphasis.
It is undoubtedly true that the  quantum effective action
of type IIB string theory on $K3\times
\wh S^1$ has a Gauss-Bonnet term
which maps to the
tree level Gauss-Bonnet term of the heterotic theory on $T^5$
after a duality transformation. Thus 
we can construct a small black hole / ring
solution of the equations of motion derived from
the quantum effective action
of the type IIB string theory on $T^5$, since the
corresponding solution exists in the heterotic theory.
However the difficulty with this is the following.
A simple analysis of the duality map tells us that the
coefficient of this term in the type IIB theory is 
proportional
to the inverse power of the radius of $\wh S^1$.\footnote{More
generally, it has been shown in appendix
\ref{sa} that all the tree
level higher derivative terms in the heterotic theory, when written
in the type IIB variables, have the property that the coefficients
of these terms have inverse powers of the radius of $\wh S^1$.
}
Thus it diverges in the limit when the radius of $\wh S^1$ goes to
zero, reflecting the fact that
in actual computation this term
arises after integrating out the  modes
carrying momentum and winding along $\wh S^1$\cite{9704145}.
Thus if
we choose to work with this quantum 
effective action we have to forget about
the original ten dimensional 
description and use an effective five dimensional
description. However the ten dimensional description
plays a non-trivial role in constructing the smooth 
classical solutions
describing the D1-D5 system in type IIB string theory. In
particular the $\wh S^1$ circle is contractible in the full
ten dimensional solution\cite{0212210}; 
hence the smooth solutions will not appear as smooth solutions
in the effective five dimensional description.
This will of course be the wrong approach to this problem,
and the correct approach will be to begin with the full ten
dimensional smooth
solution and study quantum corrections to that
solution. We expect these to be small since the ten dimensional
geometry is perfectly regular\cite{0412133}.
This shows 
that from the point of view of type IIB string theory
it will not be correct to describe these states as small black holes /
black rings
by including the quantum corrections to the five dimensional
effective action {\it computed
in the vacuum}.
On the other hand in the heterotic description we are perfectly
justified in starting with the tree level effective action and
constructing small black hole / black ring solutions 
corresponding to these states, particularly since the string coupling
remains small at the core of the solution. The extra circle that
is used to smoothen out the solutions on the type IIB side is not
visible in the perturbative heterotic string theory.

Let us now turn to another example where we choose the
compact space to be $T^5\times S^1$ on the heterotic side.
This theory has a dual description as
type IIB string theory compactified
on $K3\times  \wt S^1\times S^1$
where the configuration 
under study is mapped to a system containing $w$ 
KK monopoles associated with the circle $\wt S^1$ and
$-n$ units of momentum along $S^1$. 
First let us analyze the possibility of the existence of a small black
hole describing this system. Since in this case $w$ denotes an
NSNS sector magnetic charge and $n$ denotes an NSNS sector
electric charge, the scaling relation \refb{en1} gives
\be \label{es1}
S_{BH}( \lambda^2\, n, w) = \lambda^2 \, S_{BH}(n,w)\, .
\ee
This is incompatible with the microscopic result $4\pi\sqrt{nw}$,
showing that there cannot be any classical black hole solution associated
with these charges. 

However one can 
argue that in this case the entropy can again be reproduced
by quantization of smooth classical 
solutions\cite{9707042,0605210,0611124}.
At a generic point in the
moduli space a system of $w$ KK monopoles
is described by a smooth classical solution, -- 
the $w$-centered Taub-NUT
space.\footnote{Again if we insist on using the quantum
effective action 
we shall get a curvature squared term which can generate
a small
black hole solution with the right entropy, but the coefficient
of this term blows up as the size of $\wt S^1$ goes to zero,
indicating that these terms are generated by integrating out the
winding modes along $\wt S^1$.  Hence the inclusion of such a term
in the quantum effective action forces us to use a description in terms
of a five dimensional effective theory. In this description 
we lose the KK monopole as a smooth solution since the circle
$\wt S^1$ shrinks to zero size at the core of the KK monopole.}
This has $3w$ normalizable zero modes associated with the transverse
motion of the $w$ KK monoples, leading
to deformations parametrized by
$3w$ massless scalar fields on the $(1+1)$-dimensional
world-volume theory of this system spanned by the time 
coordinate $t$ and
the coordinate $y$ along $S^1$.\footnote{We shall take the
deformations to be independent of the coordinates of K3 in
order to preserve the BPS condition.}
Furthermore in the background of $w$
KK monopoles there are $w$ normalizable self-dual
harmonic 2-forms $\omega_m$ ($m=1,\cdots w$)\cite{ruback}. 
Thus every 2-form field $C$ of the six dimensional theory
describing type IIB string theory on K3 will give rise to
$w$ massless scalar fields $\phi_m$ on the (1+1) dimensional
world-volume spanned by $(y,t)$ coordinates via the decomposition
\be \label{edecomp}
C = \sum_{m=1}^w \phi_m(y,t)\, \omega_m\, .
\ee
Furthermore if the field strength $dC$ is 
(anti-)self-dual, the corresponding scalar fields $\phi_m$ are
(right-) left-moving on $S^1$. 
Thus the $21$ self-dual and $5$ anti-self-dual
2-form fields in type IIB string theory on K3 -- arising from
the NSNS and RR 2-form fields of the ten dimensional type IIB
string theory and the reduction of the RR 4-form field on the 2-cycles
of K3 -- give rise to deformations parametrized by
$21w$ left-moving and $5w$ right-moving scalars
along $S^1$. Thus we have altogether $24w$ left-moving and $8w$
right-moving scalars along $S^1$. There are also fermionic deformations
parametrized by $8w$ right-moving
fermion fields along $S^1$, with each KK monopole contributing 8
right-moving fermions. If we want these deformations to preserve the
supersymmetries of the undeformed background as is needed for counting
BPS states, we must 
keep the right-movers along $S^1$ in their ground state
but allow for arbitrary excitation of the 
left-movers.
Since the left-movers describe a (1+1)-dimensional conformal
field theory with total central charge $24w$, the standard application
of Cardy formula
tells us that upon quantization of these
classical deformations, 
the degeneracy of states carrying $-n$
units of momentum along $S^1$ grows as $\exp(4\pi\sqrt{nw})$, in
agreement with the microscopic answer.
Thus we see that in this case the macroscopic entropy arises 
from quantization of smooth deformations
of the $w$ KK-monopole solution. Therefore it is just as well that
in this duality frame there are no small black holes accounting for the
entropy of this system.

The above argument has been somewhat abstract, but at
least for
$w=1$ all the left-moving modes can be constructed explicitly
as
horizonless, supersymmetric,  finite  deformations of
the original Taub-NUT geometry, taking into account the effect
of gravitational backreaction.
For this we need to take the family of deformations of BMPV black
hole in Taub-NUT space constructed in\cite{0907.0593} and simply
replace the BMPV metric by flat metric. This ensures that 
quantization of
these
deformations can generate not only low lying states associated with
small fluctuations but highly excited states which contribute to the
degeneracy. 
We expect that similar construction will be possible even
for the $w\ne 1$ case -- {\it e.g.} by taking a $\ZZZ_w$
orbifold of the deformed $w=1$ solutions and then switching
on additional deformations involving twisted sector NSNS and RR
fields -- but in this case the geometric quantization
of the solutions may be more complicated as the moduli
space of multiple KK monopoles has singularities at the points
where the monopoles coincide.

An alert reader could wonder whether it is possible to get around the
scaling argument by considering small black rings or other black
objects which carry, besides the usual gauge charges, also dipole
charges\cite{0407065,0408106,0408120,0408122}. 
In this case the scaling argument will require us to also scale
the dipole charges, and we may be able to violate the scaling relations
by working in sectors with fixed dipole charges. This however does not
allow us to get around the no go results. To illustrate this let us
focus on the particular example of the D1-D5 system in type IIB string
theory on $K3\times \wh S^1$, and assume, for
the sake of argument, that there is a small black ring solution at tree
level that
carries, besides the D5-brane charge $w$ and D1-brane charge $n$,
some electric dipole charge $p$ in the NSNS sector. 
Then under the scaling
$p$ scales to $\lambda^2 p$ and the scaling relation \refb{en1}
takes the form
\be \label{edi1}
S_{BH}( \lambda\, n, \lambda\, w, \lambda^2 p) 
= \lambda^2 \, S_{BH}(n,w, p)\, .
\ee
Thus a relation of the form $S_{BH}=4\pi\sqrt{nwp}$ will be
consistent with \refb{edi1} and one could argue that the small
black ring with $p=1$ produces the desired entropy. While this
is correct, the point is that there is no reason why we should only
include in our analysis small black rings with $p=1$. In particular
the scaling argument tells us that if in the classical theory
there exists a small black
ring solution with $p=1$ and entropy
$4\pi\sqrt{nw}$ for a given $n$ and $w$, then there also exists
a small black ring with $p=\lambda^2$, D5-brane charge 
$w'=\lambda w$ and D1-brane charge $n'=\lambda n$
with entropy
$4\pi \lambda^2 \sqrt{nw}=4\pi \lambda \sqrt{n'w'}$. For
large $\lambda$ this is larger than the required answer
$4\pi\sqrt{n'w'}$, leading to a contradiction with the
microscopic results. Thus we would conclude that such small
black ring solutions do not exist in classical type IIB string
theory on $K3\times \wh S^1$.

What general lesson can we draw from this analysis? One
lesson is that in
any given duality frame, in computing the degeneracy of states using
 a macroscopic description we must include the contribution from
 both (small) black holes {\it and} smooth classical
 solutions.
 In some description
 the contribution may come solely from smooth solutions and in some
 other description it may come solely from (small) black holes and only
 after adding their contributons together we can recover the
 complete duality
 invariant answer. This is also consistent with the proposal of
 \cite{0903.1477} for a general formula 
 for the macroscopic expression for the
 degeneracy of BPS states carrying a given set of charge 
 quantum numbers $\vec q$ (including angular 
 momentum): 
 \be \label{e1x}
d_{macro}(\vec q) = \sum_s\,
\sum_{\{\vec q_i\}, \vec q_{hair}\atop \sum_{i=1}^s
\vec q_i+ \vec q_{hair}=\vec q} 
\, \left\{\prod_{i=1}^s \, d_{hor}(\vec q_i)\right\}  \,
d_{hair}(\vec q_{hair}; \{\vec q_i\})\, .
\ee
The $s$-th term on the right hand side of \refb{e1x}
represents
the contribution to the degeneracy
from an $s$-centered black hole
configuration.\footnote{In general this sum should include
 sum over all BPS black objects, including those which might
 be localized in one or more internal directions.
 For brevity we shall refer to all
 of them as black holes.}
$d_{hor}(\vec q_i)$ is the degeneracy 
associated with the horizon of
a single centered black hole (or any other black object)
carrying charge $\vec q_i$. It
is  given by the exponential of the
Wald entropy in the classical limit but more generally
by the path integral of string
fields over an
Euclidean space with the asymptotic boundary conditions set
by the attractor geometry\cite{0809.3304}.
$d_{hair}(\vec q_{hair};\{\vec q_i\})$ is the degeneracy 
associated with
the hair\cite{0901.0359,0907.0593} -- 
smooth deformations of the black hole
solution with support outside the horizon(s) --
carrying total charge $\vec q_{hair}$, 
of an $s$-centered black hole whose horizons 
carry charges $\vec q_1,\vec q_2, \cdots, \vec q_s$.
{}From this viewpoint the degeneracy of states obtained by quantizing
smooth solutions without horizon will be
represented by the $s=0$ terms in eq.\refb{e1x}, and will have to be
included in addition to the contributions from
black holes to get the complete
macroscopic result for the degeneracy. This can then be compared
with the microscopic result.

If instead of small black holes we consider 
large black holes then eq.\refb{e1x}
would lead to the folowing conclusion.
Since large black holes  have
finite area event horizon in the supergravity limit, they exist in all
duality frames, although the contribution to the horizon entropy
$d_{hor}$, obtained from path integral over string fields in the near
horizon geometry, could be different in different duality frames
after we take into account higher derivative and quantum corrections.
This would mean in particular that the right hand side of
\refb{e1x} has a non-vanishing
contribution from the $s=1$ term in all 
duality frames. Thus in no duality frame 
$d_{macro}$ will be described just by the $s=0$ term, \i.e.\
by smooth solutions without horizon. If such solutions do exist,
then their contribution must be {\it added} to the contribution
from black holes
represented by the $s\ge 1$ terms in the sum.

\sectiono{Degeneracy or Index?} \label{sindex}

In comparing the spectrum of BPS states in different descriptions
one often makes use of an appropriate index instead of absolute
degeneracy since the former is protected against quantum corrections.
As discussed in \cite{0903.1477}, one can also write down a
formula analogous to \refb{e1x} that involves the index instead
of the degeneracy. 
For simplicity we shall first 
confine our discussion to theories with
four non-compact space-time dimensions, 
and then briefly describe its
generalization to higher dimensions.
In heterotic string theory on $T^6$
the relevant index
for half BPS states is the fourth helicity trace $B_4$,
where in general $B_{2k}$ is defined as\cite{9708062,9708130}
\be \label{edefb2k}
B_{2k} = (-1)^k\, Tr\left[ (-1)^{2h} (2h)^{2k}\right] / (2k)!\, .
\ee 
Here
$h$ denotes the helicity of the state (or component of 
angular momentum along some specific direction in the rest frame)
in a fixed charge sector. For $k=0$ this will be the Witten
index. The purpose of inserting the $(2h)^{2k}$ factor in the trace
is to soak up the contribution from the fermion zero modes
associated with $4k$
broken supersymmetry generators; without this factor
the trace will vanish.
The generalization of \refb{e1x} for $B_4$ takes the 
form: 
\be \label{e2xpre}
B_{4;macro}(\vec q) = \sum_s\,
\sum_{\{\vec q_i\}, \vec q_{hair}\atop \sum_{i=1}^s
\vec q_i+ \vec q_{hair}=\vec q} 
\, \left\{\prod_{i=1}^s\, B_{0;hor}(\vec q_i)\right\}  \,
B_{4;hair}(\vec q_{hair}; \{\vec q_i\})\, ,
\ee
where $B_{0;hor}(\vec q_{hor})$ denotes the
zeroeth helicity trace -- \i.e.\ the Witten index --
of the horizon degrees of freedom and 
$B_{4;hair}(\vec q_{hair}; \{\vec q_i\})$ denotes the
fourth helicity trace of the hair degrees freedom
in given charge sectors.  
Note that in this case the
charge vector $\vec q$ no longer contains angular momentum, since
we have already summed over angular momenta for computing the
helicity trace. 
In arriving at \refb{e2xpre} we
have used the fact that the fermion zero modes associated
with broken supersymmetry are typically part of
the hair degrees of freedom
and hence we need to pick from the $(2h)^4$ factor in $B_4$
the $(2h_{hair})^4$ term in order to avoid vanishing of the
trace over the fermion zero modes. Now
in a given charge sector
$B_{0;hor}(\vec q_{hor})$ is given by 
\be \label{eb0}
B_{0;hor}(\vec q_{hor}) =\sum_{h_{hor}} \, (-1)^{2h_{hor}}
\, d_{hor}(\vec q_{hor}, h_{hor})\, ,
\ee
with $h_{hor}$ denoting the angular momentum associated
with the horizon. However 
in four dimensions supersymmetric
black hole horizons carry no angular momentum. Hence 
only the
$h_{hor}=0$ term will contribute and we shall have 
$B_{0;hor}(\vec q_{hor})=d_{hor}(\vec q_{hor},
h_{hor}=0)$.
Thus we can replace \refb{e2xpre} by
\be \label{e2x}
B_{4;macro}(\vec q) = \sum_s\,
\sum_{\{\vec q_i\}, \vec q_{hair}\atop \sum_{i=1}^s
\vec q_i+ \vec q_{hair}=\vec q} 
\, \left\{\prod_{i=1}^s\, d_{hor}(\vec q_i,h_i=0)\right\}  \,
B_{4;hair}(\vec q_{hair}; \{\vec q_i\})\, .
\ee
Eq.\refb{e2x} again shows that in
any given duality frame the contribution from the smooth solutions
represented by the $s=0$ term and small black holes
represented by the $s=1$ term must be
added. 
In fact in this case there are no contributions to the index from the
$s\ge 2$ terms, \i.e.\ multi-centered black holes. Hence the
sum of the $s=0$ and $s=1$ terms
will have to be compared with the index computed
on the microscopic side. 

For fundamental heterotic string states carrying quantum numbers
$(n,w)$ the index computed in the microscopic
description grows in the same way as the degeneracy,
\i.e.\ as $\exp(4\pi\sqrt{nw})$, up to factors involving
(inverse) powers of $nw$. On the macroscopic side there are no
smooth solutions in the heterotic description, but there are small
black holes for which 
$d_{hor}(n,w,h=0)$  grows as 
$\exp(4\pi\sqrt{nw})$.
If the only hair degrees of freedom are the fermionic zero modes 
associated with the eight broken supersymmetries, then 
$\vec q_{hair}=0$ and
$B_{4;hair}=1$. This gives $B_{4;macro}(n,w)=d_{hor}(n,w,h=0)
\sim \exp(4\pi\sqrt{nw})$.  Even if the hair modes include some
additional degrees of freedom, typically their contrbution
to $\ln B_{4;macro}$
is small compared to the horizon contribution and we still get
$B_{4;macro}(n,w)
\sim \exp(4\pi\sqrt{nw})$, {\it as long as there are no additional
fermion zero modes among the hair degrees of freedom.}
Assuming that the last condition holds, we see that 
in the heterotic description
we get a contribution to $B_{4;macro}$ of order
$\exp(4\pi\sqrt{nw})$ 
from the single black hole sector in agreement with the
microscopic result. In the type II description where the
contribution to the macroscopic entropy comes from the
smooth solutions -- {\it e.g.} as KK monopole carrying
momentum along its world-volume --
geometric quantization directly gives the
degeneracy for any angular momentum, and hence also the
index. Thus as long as the classical modes are in one to one
correspondence with the oscillation modes of the fundamental
heterotic string, we shall automatically get the correct degeneracy
and index from geometric quantization of these modes.

Dealing with the index rather than the absolute degeneracy
also throws some light on the
difficulties associated with small black holes in type II string theory
on $T^6$, representing elementary type II string states. 
If we consider type IIA/IIB 
string theory on $T^6$ and take a fundamental
type IIA/IIB 
string carrying winding charge $w$ and momentum $-n$
along one of the circles, then the microscopic degeneracy of these
states grows as $\exp(2\sqrt 2\pi\sqrt{nw})$. Thus one might ask
if these can be associated with the entropy of a small black hole.
Even though the scaling analysis \refb{e3} would tell us that 
the classical small black hole could carry entropy proportional to
$\sqrt{nw}$, explicit analysis 
has failed to find such a small black hole
essentially due to the
absence of curvature squared corrections in tree level type II string
theory.\footnote{A proposal for the near horizon theory
of these black holes has been suggested in \cite{0707.3818},
but there are no explicit solutions.}
We shall now argue that this could be a consequence of the
(generalization of the)
index formula \refb{e2x}; in fact if we had found only 
spherically symmetric
small 
black holes in type II string theory
on $T^6$, as in the case of heterotic string theory on $T^6$,
and if the only fermion zero modes associated with this solution
were the ones associated with the broken supersymmetry generators,
it would have led to a  contradiction.
The relevant states
are quarter-BPS
and the relevant index is the twelfth helicity trace $B_{12}$.
On the microscopic
side these states are obtained by keeping the
right-moving oscillators in their ground states, but considering excited
states of the left-moving oscillators.
Up to an overall normalization
the microscopic contribution to $B_{12}$ 
is given by the coefficient of $q^{nw}\, v^4$ in the expansion of
\be \label{esm2}
\prod_{k=1}^\infty \left(1-q^k e^{iv}\right)^{-1}
\left(1-q^k e^{-iv}\right)^{-1} \left(1-q^k \right)^{-6}
\left(1-q^k e^{iv/2}\right)^4 \left(1-q^k e^{-iv/2}\right)^4\, .
\ee
Physically \refb{esm2} represents the partition function of
left-moving scalars and
Green-Schwarz fermions on the fundamental string
world-sheet, with $v$ being the variable conjugate to the
helicity. 
The 8 left-handed and 8 right-handed zero modes of the
Green-Schwarz fermions are
soaked up by the eight factors of $h$ in the helicity
trace. The remaining four factors of $h$ are needed to
soak up the bose-fermi degeneracy among the states
created by the
non-zero mode oscillators, reflecting the fact that
not all the zero modes associated with the broken supersymmetries
appear as zero modes in the world-sheet theory, -- 
this is the reason why we need to
compute the coefficient of the $v^4$ term.
Collecting the order $v^4$ term in the expansion of
\refb{esm2} gives
\be \label{esm3}
\sum_{k=1}^\infty \, q^{k} \, (1 + 4 q^k + q^{2k}) (1 -q^k)^{-4}\, ,
\ee
up to an overall normalization. If $b(N)$ denotes the coefficient of 
$q^{N}$ for large $N$, then one can easily verify\cite{0507014}, 
using the
modular properties of \refb{esm2},  that for large
$N$, $\ln b(N)$ grows as $3\ln N$. 
Thus we have
\be \label{emicroind}
\ln B_{12;micro}(n,w) \simeq 3\ln (nw)\, .
\ee

Now suppose that 
we have a small spherically symmetric
quarter-BPS black hole solution,
carrying charge
quantum numbers of elementary string in type II string
theory on $T^6$, and suppose further that there are no other 
classical quarter-BPS
black hole solutions with non-zero angular momentum, 
carrying the same charge quantum numbers.
Then the argument of \cite{0903.1477} would lead to the
relation
\be \label{e3x}
B_{12;macro}(\vec q) = \sum_s\,
\sum_{\{\vec q_i\}, \vec q_{hair}\atop \sum_{i=1}^s 
\vec q_i+ \vec q_{hair}=\vec q} 
\, \left\{\prod_{i=1}^s \, d_{hor}(\vec q_i,h_i=0)\right\}  \,
B_{12;hair}(\vec q_{hair}; \{\vec q_i\})\, .
\ee
For states carrying quantum numbers of the fundamental string
there are no contributions from the $s\ge 2$ terms. On the other
hand we do not expect to have smooth classical solutions
describing fundamental type II string, and so the $s=0$ term also
does not contribute. Thus we focus on the contribution from the
$s=1$ term, \i.e.\ single centered black hole solutions. 
If the only fermion zero modes associated
with the hair are those associated with the twelve broken 
supersymmetry
generators, then $B_{12}$ associated with the hair 
carrying $\vec q_{hair}=0$ is non-zero and 
the helicity trace 
\refb{e3x}
receives a contribution proportional to
$d_{hor}(n,w,h=0)$. Now
based on general scaling arguments given earlier 
we shall have
\cite{9712150}
\be \label{esm1}
\ln d_{hor}(n,w,h=0) = C\, \sqrt{nw}\, ,
\ee
for some constant $C$.
This would give
\be \label{eloghel}
\ln B_{12;macro}(n,w) \simeq C\, \sqrt{nw}\, .
\ee

This is in clear contradiction to \refb{emicroind}.
There seem to be two natural possibilities:
1) there is no small black hole describing the
fundamental type II
string; 
2) there is  a spherically symmetric small black hole solution,
but the hair degrees of freedom associated with this black hole
carry additional fermionic zero
modes besides the ones associated with
the broken supersymmetry generators. In this case quantization of 
these additional fermion zero modes will make $B_{12;hair}$
vanish and hence this small black hole will not contribute to
$B_{12;macro}$.

We can gain further insight by taking an asymmetric 
orbifold of type IIA/IIB string theory on $T^6$ by  
$(-1)^{F_L}$ accompanied
by a half shift along one of the circles $S^1$. This gives
an
$\NN=4$ supersymmetric string theory for which the helicity
trace $B_4$
associated with the elementary string 
states grow as
$\exp(2\sqrt 2\pi\sqrt{nw})$\cite{0504005}.
Thus supersymmetric small
black holes must show up on the
macroscopic side. 
Furthermore, the scaling argument of \S\ref{small} tells us
that in order that the logarithm of the
index computed from this black
hole is of order $C\sqrt{nw}$ it must arise at string tree 
level\cite{9504147,9712150}. As a result it must also exist in
type II string theory on $T^6$, since the tree level effective action
is not affected by compactification.
The microscopic contribution to the helicity
trace in the orbifold theory
however has the strange property that it oscillates between
positive and negative values depending on
whether the momentum along $S^1$ is even or odd. Such a
behaviour of the index cannot be reproduced by a 
spherically symmetric extremal
black hole 
whose only hair degrees of freedom are the fermion zero modes
associated with the broken supersymmetry generators; such a
system will
always give a positive contribution $d_{hor}(n,w;h=0)$ 
to the index.
Instead we need to look for new classes of supersymmetric
extremal small black holes carrying non-zero angular momentum
correlated with the momentum along $S^1$.

Combining all these informations we  shall now
propose a
possible scenario, but we do not claim that this is the only
possible resolution. In this scenario
type II string theory on $T^6$ has a
quarter BPS
small black hole carrying the same charges as the elementary string
states, but besides the twelve fermion zero modes associated with
broken supersymmetry generators, the world-volume degrees of
freedom of the black hole also contain a massless
(1+1) dimensional fermion field on $S^1$
arising out of R-NS sector 
fields.\footnote{For black holes whose near horizon
 geometry contains an $AdS_2$ factor, we expect a clean
 separation between the hair and the horizon degrees of freedom
 due to the infinite $AdS_2$ throat. 
 However it is not necessary that the proposed small black hole
 of type IIB string theory should have an $AdS_2$ factor, and hence
 there may not be a separation between the hair and horizon degrees
 of freedom. For this reason it may not be meaningful to ask whether 
 this
 fermionic field lives on the hair or inside the horizon.}
Thus these modes are odd under $(-1)^{F_L}$. 
Due to the  additional
zero modes from this fermion field
this small black hole will not contribute to the
index $B_{12}$ in type IIB string theory on $T^6$. However
since these zero modes are projected out in the orbifold theory,
the black hole in the orbifold theory will have non-vanishing index.
Furthermore even though the zero modes are projected out,
the fermionic modes  carrying odd
momentum along $S^1$ survive the projection since they are even
under the combined operation of a half 
shift along $S^1$ and
$(-1)^{F_L}$. Since in order to get a state with odd momentum
along $S^1$ we need to excite odd number of modes of this fermion,
they will be fermionic states and hence will give a negative contribution
to the index. This would automatically explain why in the
orbifold theory $B_{4;macro}$ oscillates between 
positive and negative
values as we change the momentum along $S^1$.

This unusual property of these black holes, -- namely presence of
additional fermionic  modes on the world-volume, 
-- could be the reason that
these are not seen as solutions to the effective action containing just
the `F-terms'; additional higher derivative D-terms 
may be needed for
their existence. Such terms are undoubtedly present in the
effective action of type II string theory compactified on $T^6$,
but we have much less control on these terms to be able to analyze
small black hole solutions in their presence.

Before concluding this section we shall briefly describe the
generalization of this analysis to higher dimensions.
In four dimensions the
information about the angular
momentum carried by the individual states is lost in the
index since we have to sum over
all states carrying different angular momenta in defining the
index. In five and higher 
dimensions one can do somewhat better.
For definiteness we
shall focus on the five dimensional theory
obtained by compactifying heterotic string theory on $T^5$.
In this case the rotation group is $SO(4)\simeq SU(2)\times SU(2)$
and we can use a pair of qunatum numbers $(h_1, h_2)$,
labelling the $J^3$ quantum numbers of the two $SU(2)$'s,
to characterize a state. We can now define an index
\be \label{e5ind}
I_{2k}(\vec q, h_1) = {(-1)^{k}\over (2k)!}\,
Tr_{\vec q, h_1} ((-1)^{2h_1+2h_2} \, 
(2h_2)^{2k})\, .
\ee
Note that the trace is taken over a fixed charge and fixed $h_1$ sector.
In defining the index we have broken the symmetry between the two
angular momenta; which one we choose to keep fixed is a matter
of convention. The index $I_2$
for fixed quantum numbers $(n,w,h_1)$ receives contribution 
only from
the half BPS states of the heterotic string. 
To see this we note that a half BPS state will break
eight supersymmetries, and the associated fermion zero modes will
contain four modes with $(h_1, h_2)=(0,\pm 1/2)$ and four modes with
$(h_1,h_2)=(\pm 1/2,0)$. The $(2h_2)^2$ factor 
is needed to soak up the fermion
zero modes with quantum numbers $(0,\pm 1/2)$. 
On the other hand since
the trace is taken over a fixed $h_1$ sector the fermion zero modes
carrying quantum numbers $(\pm 1/2,0)$ do not give a vanishing
result. For a non-BPS state the number of fermion zero modes double
in each sector and the trace over the eight fermion zero modes carrying
quantum numbers $(0,\pm 1/2)$ will make the result vanish.

The analog of the formula \refb{e2xpre}
for the macroscopic index
now takes the form:
\be \label{e2x5}
I_{2;macro}(\vec q, h_1) = \sum_s\,
\sum_{\{\vec q_i\}, \vec q_{hair}\atop \sum_{i=1}^s
\vec q_i+ \vec q_{hair}=\vec q, \sum_{i=1}^s
  h_{1i}+   h_{1;hair}=  h_1} 
\, \left\{\prod_{i=1}^s\, I_{0;hor}(\vec q_i; h_{1i})\right\}  \,
I_{2;hair}(\vec q_{hair}; \{\vec q_i\}; h_{1,hair})\, ,
\ee
again taking into account the fact that the fermion zero modes associated
with broken supersymmetry generators are part of the hair degrees
of freedom. In the $s=1$ sector of the heterotic description,
\i.e.\ the sector containing a single small `black object',
as long as the only fermion zero modes associated with
the hair degrees of freedom are the ones associated with the broken
supersymmetry generators, $I_{2;hair}$ 
will give a
finite contribution. 
In particular if the only hair degrees of freedom are the fermion
zero modes associated with the broken supersymmetry generators
then $\vec q_{hair}=0$ and $I_{2;hair}=-1,2,-1$ for
$h_{1,hair}=-{1\over 2}, 0, {1\over 2}$.\footnote{These
numbers arise from
the quantization of the four
fermion zero modes carrying $(h_1,h_2)=(\pm 1/2, 0)$.}
This gives
\be \label{ei2rel}
I_{2;macro}(\vec q, h_1)=2I_{0;hor}(\vec q, h_1)
- I_{0;hor}(\vec q, h_1-{1\over 2}) - I_{0;hor}
(\vec q, h_1+{1\over 2})\simeq -{1\over 4}\, {\p^2 I_{0;hor}(\vec q, h_1)
\over \p h_1^2}\, .
\ee
On the other hand
$I_{0;hor}(n, w, h_1)$ can be equated to 
$\sum_{h_{2}} (-1)^{2h_{1}+2h_{2}} 
d_{hor}(n,w,h_1,h_2)$. As discussed in
\S\ref{small},
in this case the contribution to $d_{hor}(n,w,h_1,h_2)$
comes from a small black ring rotating in a plane,
\i.e.\ carrying  $|h_1|=|h_2|$,
and the entropy is given by 
$4\pi\sqrt{nw - |2h_1|}$.
This would
give 
\be \label{egi1}
I_{0;hor}(n, w, h_1) \sim \exp[4\pi\sqrt{nw - |2h_1|}]\, .
\ee 
When we substitute this in \refb{ei2rel} the derivatives with
respect to $h_1$ bring down inverse powers of charges,
but do  not curb the exponential growth.
As a result we have
\be \label{egi1rep}
I_{2;macro}(n, w, h_1) \sim \exp[4\pi\sqrt{nw - |2h_1|}]\, ,
\ee
where $\sim$ denotes equality up to factors involving (negative)
powers of $n$, $w$ and $h_1$.
As in the case of four dimensional small black holes, this result
continues to hold even if the hair modes include some additional
degrees of freedom, as long as they do not have any additional
fermion zero modes. 

On the other hand the microscopic computation of this index 
can be performed using  the
known spectrum of elementary heterotic string states.
The BPS states correspond to states of the
fundamental string  for which the right-movers
are kept in their ground state.
The  index $I_2(n,w,h_1)$ is given by the coefficient of 
$q^{nw} z^{2h_1}$ term in
the expansion of:
\be \label{emic2}
q^{-1} \prod_{k=1}^\infty (1 -q^k)^{-20} \, (1-q^k z)^{-2} 
(1-q^k z^{-1})^{-2}\, (i z^{1/2} - i z^{-1/2})^2\, .
\ee
Here the $(1 -q^k)^{-20}$ term is the 
contribution from the left-moving
bosonic oscillators along the internal directions, and the 
$(1-q^k z)^{-2} 
(1-q^k z^{-1})^{-2}$ term represents contribution from the 
left-moving bosonic oscillators along the four 
non-compact transverse directions.
The $(i z^{1/2} - i z^{-1/2})^2$ term is the contribution from the
four fermion zero modes carrying $(h_1,h_2)=(\pm 1/2, 0)$.
Using a slight variation
of the method developed in \cite{9405117} one finds that
for large $nw$ and $h_1$ the index grows as
\be \label{egrow}
\exp[4\pi\sqrt{nw - |2h_1|}]\, ,
\ee
up to factors involving powers of the charges. This is
in agreement with the macroscopic result \refb{egi1rep}.

In the type IIB description where these states are described as smooth
classical solutions one needs to calculate the index by 
geometric quantization of the classical solutions with the 
the extra factor of $((-1)^{2h_1+2h_2} \, 
(2h_2)^{2})$ inserted. Since these classical solutions are in
one to one correspondence to classical oscillations
of the dual heterotic string, we expect the
index to be reproduced correctly after we quantize this system.

\sectiono{How Does a BPS State Lose Information About
Unprotected Quantities?}
\label{sprofile}

In \S\ref{sindex} we have discussed
comparison of the
results of microscopic and macroscopic computation 
based on protected quantities like the helicity trace indices.
One might wonder if we can compare more; e.g. the wave-functions
computed in different descriptions, which
in the classical limit gives information about the profile of the
solutions.

We cannot prove that this is impossible, but there is no 
known index
theorem that protects such information. In what follows we shall
suggest a possible mechanism by which a BPS state could lose
such information by drawing analogy to chiral fermions -- an analogy
often cited in motivating the index theorem for BPS states.
As is well known,  the
difference between the number of left and right chiral fermions is 
protected by an index theorem. We shall 
now illustrate how a chiral fermion
can lose information about its wave-function and hence various
attributes by mixing with non-chiral fermions while preserving the
net number of chiral fermions in accordance with index theorem.
Let us consider a theory with a single right-chiral fermion
$\psi_R$ and $N$ non-chiral fermions $(\chi_{iL}, \chi_{iR})$
for $1\le i\le N$ with standard kinetic terms
and let us suppose that in the unperturbed theory
the mass term has the form
\be \label{emass1}
M \sum_{i=1}^N \, \bar\chi_{iL} \chi_{iR} + c.c.\, .
\ee
In this theory $\psi_R$ 
represents the massless fermion, -- the analog of
the BPS state. Now add a perturbation of the form
\be \label{emass2}
\eps\, \sum_{i=1}^N \, a_i  \, \bar \chi_{iL}\, \psi_{R}  + c.c.\, ,
\ee
where $\eps$ is a small mass
parameter and the $a_i$'s are finite dimensionless constants.
We can make a unitary rotation
 $\wt\chi_{iL}=U_{ij} \chi_{jL}$, 
$\wt\chi_{iR}=U_{ij} \chi_{jR}$, $U^\dagger U=1$,
 such that
\be \label{emass3}
\wt \chi_{1L} = |\vec a|^{-1} \, \sum_{i=1}^N \, a_i^*  \,  \chi_{iL}\, .
\ee
In this basis the mass term takes the form
\be \label{emass4}
\eps \, |\vec a|\, \bar {\wt\chi}_{1L}\, \psi_{R} +
M \sum_{i=1}^N \, \bar{\wt\chi}_{iL} \wt\chi_{iR}  + c.c.\, .
\ee
Finally we define
\be \label{emass5}
\wh\chi_R = (\eps |\vec a| \psi_R + M \wt\chi_{1R}) / \sqrt{
\eps^2 \vec a^2 + M^2}, \qquad  \wh\psi_R=
(\eps |\vec a| \wt\chi_{1R} - M  \psi_R) / \sqrt{
\eps^2 \vec a^2 + M^2}\,
\ee
so that $\wh\psi_R$, $\wh \chi_R$, $\wt \chi_{1L}$ 
and $\wt \chi_{iR}$,
$\wt\chi_{iL}$ for
$2\le i\le N$ have standard kinetic term. In this basis
the mass term is given by
\be \label{emass6}
\sqrt{
\eps^2 \vec a^2 + M^2} \, \bar{\wt\chi}_{1L} \wh \chi_{R}
+ M \sum_{i=2}^N \, \bar{\wt\chi}_{iL} \wt\chi_{iR}  + c.c.\, .
\ee
Thus in the final theory $\wh\psi_{R}$ represents the chiral fermion.
Since $\vec a^2 \sim N$, we see from \refb{emass5} it is made
predominantly out of $\wt \chi_{1R}$ if $N\eps^2 >> M^2$. Thus
even with small mixing the wave-function of the chiral fermion can
change completely if there are a large number of states it could mix
with.

To make this simple result look more dramatic we could imagine
that the theory under consideration comes from a brane-world
model in which
all the fermions live on some space filling branes, with
$\psi_R$ living on brane $A$ and the $\chi_i$'s living on brane B,
and suppose further that the branes A and B are separated by a finite
distance along some compact direction. Then in the unperturbed theory
the chiral fermion lives on the brane A. But once we switch on the
deformation we see that the chiral fermion lives 
predominantly on the brane B. Thus it
chooses to live not on the brane in which the original chiral fermion lived,
but where most of the (non-chiral) fermions lived!

If the same mechanism operates on
BPS states then this will mean that even for
a small deformation the wave-function of a BPS state can change
considerably by mixing with those of
non-BPS states if there are a large number
of non-BPS states carrying the same charges. This is precisely the
situation in string theory where for fixed set of charges, and say
within one string units of energy, the number
of non-BPS states is exponentially larger than the number
of BPS states. Thus 
the information about the wave-function of the BPS state may be
wiped out by such mixings.
(See \cite{0701122} for a somewhat related discussion.)

\bigskip

\noindent {\bf Acknowledgement:}
I would like to thank Borun Chowdhury, 
Rajesh Gopakumar, Dileep
Jatkar, Samir Mathur, 
Shiraz Minwalla, Joan Simon, Yogesh Srivastava and
Edward Witten for useful
discussions. I also thank Dileep
Jatkar, Samir Mathur, 
Joan Simon and Yogesh Srivastava for their comments on various
earlier versions of the manuscript.
I would like to acknowledge the hospitality of 
LPTHE, Paris
where part of this work was performed.
This work
was supported by the project 11-R\& D-HRI-5.02-0304 and the J.C.Bose
fellowship of the Department of Science and Technology,
India.

\appendix

\sectiono{Duality Transformation from Heterotic to 
Type II Description}
\label{sa}

In this appendix we shall analyze the fate of
the tree level effective action of heterotic string theory
under the duality map that takes it to the type IIB string theory.
We begin with heterotic string theory on $T^4$ 
which is dual
to type IIA string theory on 
K3\cite{9410167,9501030,9503124,9504027,9504047}. 
In the supergravity
approximation where we keep only two derivative terms, the
two actions are mapped to each other under this duality
transformation. The heterotic metric
$G^{(H)}$ is related to the type IIA metric $G^{(A)}$ and the
type IIA dilaton $\phi_{A}$ by the 
relation
\be \label{esa1}
G^{(H)} = e^{-2\phi_{A}} \, G^{(A)}\, .
\ee
Consider now a term in the tree level heterotic
string theory carrying $2n+2$ derivatives. This will carry $n$
extra powers of the inverse metric of heterotic string theory
compared to the two derivative terms, and hence,
when  expressed in the type IIA variables, will carry
an extra power of $e^{2n\phi_{A}}$ compared to the
supergravity action. In
particular if we consider only the NSNS sector fields of IIA
the term in the type IIA action will carry a factor of
$e^{2(n-1)\phi_{A}}$ since the supergravity action carries
a factor of $e^{-2\phi_{A}}$.

Now consider compactifying this theory on a circle $S^1$.
On the heterotic side the state of interest is the one that carries
fundamental string winding charge
and momentum along $S^1$.
On the type IIA side this corresponds to NS 5-branes wrapped
on $K3\times S^1$ and momentum along $S^1$. Since
both charges are in the NSNS sector we are justified in
working with the effective action in the NSNS sector. The effective
five dimensional action on the type IIA side, coming from the
tree level $2n+2$ derivative terms in the heterotic string theory,
will then contain a factor of 
\be \label{esa1.5}
e^{2(n-1)\phi_{A}}R_{A}\, ,
\ee
where $R_{A}$ is the radius of $S^1$ in the type IIA metric.
We now make  a T-duality transformation
along $S^1$ to go to the type IIB
description. The dilaton $\phi_{B}$ and the radius $R_{B}$ of
the dual circle $\wh S^1$
will be related to $\phi_{A}$ and $R_{A}$
via the relations
\be \label{esa2}
R_{A} = 1/R_{B}, \qquad e^{\phi_{A}} = 
e^{\phi_{B}} / R_{B}\, .
\ee
Using \refb{esa1.5}, \refb{esa2}
we see that the effective action in the type IIB string theory, coming
from the tree level $2n+2$ derivative term of the heterotic
string theory will contain a factor of 
\be \label{esa3}
e^{2(n-1)\phi_{B}}(R_{B})^{1-2n}\, .
\ee
In the IIB description the original fundamental heterotic string
carrying winding and momentum along $S^1$
gets mapped to an NS 5-brane wrapped on $K3\times \wh S^1$
and a fundamental string wrapped on $\wh S^1$. A further 
strong-weak coupling duality transformation of the ten dimensional
type IIB string theory will take this to a D1-D5 system, but this
does not change the power of $R_{B}$ in the effective action.
Thus from \refb{esa3} we see that any higher derivative 
($n\ge 1)$ term in the tree level
heterotic string effective action, when represented in the type IIB
description, carries an inverse power of the radius of the circle
measured in the type IIB metric.


\end{document}